# Criticism of Knapsack Encryption Scheme

Sattar J Aboud

**Abstract**— In this paper, we analyze a knapsack schemes. The one is suggested by Su, which is relied on a new method entitled permutation combination method. We demonstrate that this permutation method is useless to the security of the scheme. Since the special super increasing construction, we can break this scheme employ the algorithm provided by Shamir scheme. Finally, we provide an enhanced version of Su scheme to avoid these attacks.

—————————— ◆ ——————————

## 1 INTRODUCTION

THE knapsack problem is a well-known NP-complete problem. This problem is declared by provided positive integer elements $a_1, a_2, ..., a_n$ and $s$, if there is a subset of $a_i$ that sums to $s$. That is equal to determine if there are $m_1, ..., m_n$ where $s = \sum_{i=1}^{m} m_i * a_i$ such that $m_i \in (0,1)$ where $1 \leq i \leq n$ (1.1)

The density of the knapsack is determined $d = n / \log_2 A$, such that $A = \max_{1 \leq i \leq n} * a_i$. The difficulty of knapsack system motivated many public key schemes in the eighties, following the influential work of Merkle and Hellman [1]. Though the underlying problem is NP-complete, it has astonishingly been broken by Shamir [2] since of the special construction of the secret key. Later, many variants have been demonstrated insecure for practical applications by lattice reduction methods [3], also, Baocang Wang, Hui Liu and Yupu Hu [4] reported that common knapsack can be solved under the actuality of an Oracle when the density $d < 0.646$, which is widespread by Coster [5, 6] to $0.9408$. In fact, we can employ the celebrated LLL method [7, 8] instead of the Oracle in practical parameters. But, most of suggested knapsack systems have been broken except Okamoto, Tanaka and Uchiyama, quantum knapsack scheme [9]. Either by special construction for example Merkle and Hellman scheme and Chor and Rivest scheme [10]. To make sure of the challenge of low-density attack, it is proposed that the density $d$ should be bigger than $0.9408$ [5, 11]. Lately, some knapsack schemes with high density have suggested [12, 13].

In this paper, we study a Su knapsack scheme [14]. This scheme considered a new permutation method entitled permutation combination method, by developing this method to avoid the low-density attack. We demonstrate that the permutation method is useless to avoid the low-density attack and the density of knapsack sequence is lesser than $0.9408$. But, when we employ the low-density scheme, the dimension of the lattice is 1025, which is very large to employ LLL method [7]. Since of the special super increasing construction, we can get equal secret keys by Shamir scheme. We demonstrate that one can decrypt the message if he can solve a knapsack scheme such as the Merkle and Hellman scheme. In this paper, we study Shamir attack on the basic Merkle and Hellman knapsack scheme. Initially, we provide a concise description of the pioneering Merkle and Hellman knapsack scheme. The sender selects a super increasing vector $B = (b_1, b_2, ..., b_n)$, that is $b_j > \sum_{i=1}^{j-1} b_i$ such that $2 \leq j \leq n$ and two positive integers $w$ and $p$, with $p > \sum_{i=1}^{n} b_i$ where $\gcd(w, p) = 1$ calculates $a_i' \equiv b_i * w \bmod p$
where $0 < a_i' < p$ (2.1)
Then chooses a permutation $\pi$ of $(1, 2, ..., n)$ and determines $a_i = a_{\pi(i)}'$ where $1 \leq i \leq n$

## 2 KNAPSACK PUBLIC KEY SCHEME

The public key is the vector of $n$ positive integers $a_1, a_2, ..., a_n$ and the secret key is the super increasing vector $B$, the integers $w, p$ and the permutation $\pi$. Typically, the length of every $b_i$ is $n + i$ bits, for $1 \leq i \leq n$, the length of $p$ is $2 * n + 1$ bits. In the original Merkle and Hellman scheme $n = 100$ the message $m = (x_1, ..., x_n)$ is encrypted as

$c = \sum_{i=1}^{n} x_i * a_i$ and the receiver calculates

$$m \equiv c * w^{-1} \bmod p$$
$$\equiv \sum_{i=1}^{n} x_i * a_i * w^{-1} \bmod p$$
$$\equiv \sum_{i=1}^{n} x_i * a_{\pi(i)}' * w^{-1} \bmod p$$

————————————————
Sattar Aboud is the Information Technology Advisor, with the Iraqi Council of Representatives

$$\equiv \sum_{i=1}^{n} x_i * b_{\pi(i)} \mod p$$

Because $p > \sum_{i=1}^{n} b_i$, we can get that $c = \sum_{i=1}^{n} x_i * b_{\pi(i)}$, the formula is simple to solve because the $b_i$ form a super increasing vector. Let $u = w^{-1} \mod p$, where $0 < u < m$, from formula (2.1) $a_i \equiv b_{\pi(i)} * w \mod p$ We have $b_{\pi(i)} \equiv a_i * u \mod p$. This means that there is certain integer $k_i$ where $a_i * u - k_i * p = b_{\pi(i)}$, hence $\frac{u}{p} - \frac{k_i}{a_i} = \frac{b_{\pi(i)}}{a_i * p}$. The cryptanalyst does not know $u, p, \pi, k_i$, and $b_i$. But he can obtain the length of $a_i$ where $1 \le i \le n$ and $u$ are the same as $p$'s with $b_i \le 2^{n+i}$. The first five factors of the super vector satisfy that $b_i \le 2^{n+i}$ where $1 \le i \le 5$. Let $i_j = \pi^{-1}(j)$, then we get $\left| \frac{u}{p} - \frac{k_{i_j}}{a_{i_j}} \right| \le \frac{2^{n+5}}{2^{4*n+2}}$ where $1 \le j \le 5$ (2.2) and subtract $j = 1$ term from the others, we have $\left| \frac{k_{i_j}}{a_{i_j}} - \frac{k_{i_1}}{a_{i_1}} \right| \le 2^{-3*n+4}$ where $2 \le j \le 5$. This implies that $\left| k_{i_j} * a_{i_1} - k_{i_1} * a_{i_j} \right| \le 2^{n+6}$ where $2 \le j \le 5$ (2.3)

Formula (2.3) demonstrate how unusual $a_{i_j}$ and $k_{i_j}$ are. Finally, the length of every of them is $2*n$ bits, thus the length of $k_{i_j} * a_{i_1}$ and $k_{i_1} * a_{i_j}$ is $4*n$ bits. However the length of the difference of two such terms to be $n + 6$ bits, which requires certain special construction. Shamir core contribution was to observe that $k_{i_j}$ can be found in polynomial time by invoking Lenstra theorem that the integer programming problem in a determined number of variables can be solved in polynomial time [2]. But, the cryptanalyst must invoke $n$ times of Lenstra integer programming method since he does not know the permutation $\pi$. This produces the $k_{i_j}$ where $1 \le j \le 5$. Once $k_{i_j}$ are found, one gets an estimate to $u/p$ from the formula (2.2) and builds a pair $(u', p')$ with $u'/p'$ close to $u/p$ where the weights get by $c_i \equiv a_i * u' \mod p'$ where $0 < c_i < p'$ such that $1 \le i \le n$ form a super increasing vector. Thus one can obtain an equal private key in polynomial time.

## 3 LOW DENSITY ATTACKS

In this section, we will present the low density attack suggested by Coster [5] which is the modification of Lagarias and Odlyzko attack [4]. The knapsack system is declared as follows: For provided positive integers $a_1,...,a_n$ and $s$ obtain variables $e_1,...,e_n$ with $e_i \in (0,1)$ where $\sum_{i=1}^{n} e_i * a_i = s$. Determine the sequences $b_1,...,b_{n+1}$ as follow:

$b_i = (1,0,...,0, n*a_1)$
$b_2 = (0,1,...,0, n*a_2)$
$\vdots$
$b_n = (0,0,...,1, n*a_n)$
$b_{n+1} = (\frac{1}{2}, \frac{1}{2},..., \frac{1}{2}, n*s)$

Such that $n > \frac{1}{2}\sqrt{n}$. Let $L$ be the lattice spanned by the sequences $b_1,...,b_{n+1}$. Observe that the sequence $e = (e'_1,...,e'_n, 0) \in L$ such that $e^e_i = e_i - \frac{1}{2}$ Coster demonstrated that if the density $d < 0.9408$, the sequence $e$ is shortest sequence in $L$. Thus one can obtain $e$ when there is a SVP oracle. In reality, we usually employ LLL method [6] or other lattice bases reduction method [8] instead of the SVP oracle. But, it is thought that if the dimension is big sufficient for example bigger than 400[15], one cannot obtain the shortest sequence for random lattice.

## 4 DESCRIPTION OF SU SCHEME

We will depict the Su scheme [14] which is relied on the use of the elliptic curve logarithm problem and knapsack problem as follows: In first stage, the user chooses the elliptic curve domain elements.

- A curve $y^2 = x^3 + a*x + b$ over $F_p$ such that $4*a^3 + 27*b^2 \ne 0$.
- A point $h = (x_0, y_0)$ where $ord(h) > p$ such that $p$ is a prime number.
- A super increasing vector $v' = (a_1, a_2,...,a_n)$, where $a_1 \approx 2^n, a_i > \sum_{j=1}^{i-1} a_j$ such that $2 \le i \le n$ where $a_n = 2^{2*n}$ such that $\sum_{i=1}^{n} a_i < p$.
- Determine the vector $v = a_i * h$ where $1 \le i \le n$
- Choose a permutation $\pi$ of $(1,2,...,n)$
- Choose integer number $e, d$ where $\gcd(p, e) = 1$, such that $e * d \equiv 1 \mod p$
- Suppose $w = (s_i \equiv (e * a_{\pi(i)} \mod p)) * h$ where $1 \le i \le n$
- Suppose $f = (t_i \equiv e * a_{\pi(i)} \mod p)$ where $1 \le i \le n$
- The public key is $v, w, h, p$. The private key is $e, d, \pi$ and the super increasing vector $v'$.

**Encryption**
- Entity $B$ must do the following:

- Encrypts the original message $M = (m_1,...,m_n)$ to be passed as $x-y$ points $o_{m_i}$. Thus the message becomes $M = (o_{m_1},...,o_{m_n})$
- Selects a binary message $z = (x_1, x_2,...,x_n)$
- Selects an arbitrary integer number $r$ where $r < p$.
- If $x_i = 1$ determine $C_{m_i} = (r*h, o_{m_j} + r_{s_i})$, else, add confusing information to $o_{m_i}$. The encrypted message is $c = C_{m_1} \| C_{m_2} \| ... C_{m_n} \| \sum_{i=1}^{n} x_i * t_i \mod p$
- Send c to entity $A$

**Decryption**
Entity $A$ must do the following:

- Finds $g = e*C_{m_1} \| e*C_{m_2} \| ..., e*C_{m_n} \| d*\sum x_i * t_i \mod p$

  such that $e*C_{m_1} = o_{m_1} + r*s_i - e*r*a_{\pi(i)}*h$
  $= o_{m_1} + r*e*a_{\pi(i)}*h = o_{m_1}$ (5.1)

- Because $d*\sum_{i=1}^{n} x_i * t_i \mod p \equiv \sum x_i * a_{\pi(i)} \mod p = \sum_{i=1}^{n} x_i * a_{\pi(i)}$

- Entity $A$ can get $z = (x_1,...,x_n)$ by solving the super increasing knapsack and gets $o_{m_i}$ from formula (5.1) for $x_i = 1$.

- Decrypts $o_{m_1}$ to obtain the message $m$

**4.1 Attack of Su Scheme**

Notice that the vector $z = (x_1,...,x_n)$ is influential. From this sequence, Entity $A$ can obtain the right $C_{m_i}$ from the confusing information. Initially, we provide a known the sequence $z$ attack and this attack will be employed in the following. Known $z$ Attack when we know the sequence $z = (x_1, x_2,...,x_n)$, from formula (5.1), we know $e*C_{m_j} = o_{m_i} + r*s_i - e*r*a_{\pi(i)}*h$. We observe that $t_i = e*a_{\pi(i)}$. Then we can get that $e*C_{m_j} = o_{m_i} + r*s_i - e*r*a_{\pi(i)}*h = o_{m_i} + r*s_i - t_i*r*h$ unfortunately, the set $f$ is the public key in the scheme. Thus we can decrypt the message if we know the set $z$. Find the set $z$. To find the set $z$, we should solve a Merkle and Hellman algorithm. Now we have a super increasing vector $v' = (a_1,...,a_n)$ satisfies $a_1 \approx 2^n$ such that $a_i > \sum_{j=1}^{i-1} a_j$ where $2 \le i \le n$ such that $a_n \approx 2^{2*n}$ where $\sum_{i=1}^{n} < p$ and a permutation $\pi$ and the public key is the vector $f = t_i \equiv e*a_{\pi(i)} \mod p$ where $1 \le i \le n$. The private key is $v', e, \pi$ while the density $d = \frac{n}{\log_2 o} \le \frac{n}{2*n} = \frac{1}{2}$. If $n$ is moderate, we can employ low density attack [4, 5] to obtain the set $z$. When $n$ is big, we can still obtain the set $z$ by employing Shamir scheme. As a result, we can determine the message after performing a known $z$ attack. Lastly, we provide an enhanced version of this scheme to prevent the known $z$ attack. In Su paper, if $x_i = 1$, determine $C_{m_i} = (r*h, *o_{m_j} + r*s_i)$. When we rearrange $C_{m_i} = (r*a_{\pi(i)}*h, o_{m_j} + r*s_i)$, this enhanced scheme will be safer than the old scheme. Though one can obtain the set $z$ cannot get back $o_{m_i}$, since $e$ is the private key and it is difficult to calculate the elliptic curve discrete logarithm problem.

## 5 CONCLUSION

In this paper, we study a new scheme relied on knapsack system. The security level of the encryption scheme is overestimated. We can break Su scheme employ Shamir scheme and decrypt the message of the scheme.

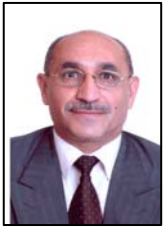

**Sattar J Aboud** is a Professor and advisor for Science and Technology at Iraqi Council of Representatives. He received his education from United Kingdom. Dr. Aboud has served his profession in many universities and he awarded the Quality Assurance Certificate of Philadelphia University, Faculty of Information Technology in 2002. Also, he awarded the Medal of Iraqi Council of Representatives for his conducting the first international conference of Iraqi Experts in 2008. His research interests include the areas of both symmetric and asymmetric cryptography.